\documentstyle[12pt, amsfonts, hyperref]{article}
\def\({\left(}
\def\){\right)}
\def\[{\left[}
\def\]{\right]}
\def\vect#1{\skew{-1}{\mathaccent"017E}{#1}}
\def\vp{\vect p}
\def\vx{\vect x}

\def\be{\begin{equation}}
\def\e{\end{equation}}
\def\ee{\end{equation}}
\begin{document}

\title{On the Casimir energy for scalar fields with bulk inhomogeneities}

\author{I.V. Fialkovsky${}^{a}$\thanks{%
e-mail: ignat.fialk@paloma.spbu.ru},
V.N. Markov${}^{b}$\thanks{%
e-mail: markov@thd.pnpi.spb.ru},
Yu.M. Pis'mak${}^{a}$\thanks{%
e-mail: pismak@JP7821.spb.edu}
\\
$^{a)}${\small St Petersburg State University, St Petersburg, Russia}
\\
$^{b)}${\small St Petersburg Nuclear Physics Institute, Gatchina, St Petersburg, Russia}}

\maketitle

\begin{abstract}
We study the field theoretical model of a real scalar field
in presence of spacial inhomogeneity in
form of a finite width mirror (material layer). The interaction of the
scalar field with the defect is described with position-dependent
mass term. We calculate the propagator of the theory, the
Casimir energy and the pressure on the boundaries of the layer. We
discuss the renormalization procedure for the model in dimensional
regularization.
\end{abstract}

\section{Introduction}
Quantum Field Theory (QFT) was developed in the middle of the last
century as a theory of interaction of elementary particles in
otherwise empty, homogenous infinite space-time \cite{BogShir}. On the other
hand, from the very beginning it was clear that presence of
boundaries, non-zero curvature or nontrivial topology of the
space-time manyfold should influence the spectrum and dynamics of
the excited states of the model as well as the properties of the
ground state (vacuum).

The first quantitative description of such changes in the
vacuum properties was made by H. Casimir in 1948. He predicted
\cite{Casimir'48} macroscopical attractive force between two
uncharged conducting plates placed in vacuum. The force
appears due to the influence of the boundary conditions on the
electromagnetic quantum vacuum fluctuations. Nowadays the Casimir
effect is verified by experiments with the precision of
$0.5$\% (see \cite{Klimchitskaya 05} for a review).

The properties of the vacuum fluctuations in curved spaces,
investigation of scalar field models
with various boundary conditions and their application to the
description of real electromagnetic effects  were actively
studied through the last decades, see discussion and references in
\cite{Klimchitskaya 05}, \cite{Milton-OBZOR'04}.

However, it was well understood that boundary conditions must be
considered just as an approximate description of complex interaction
of quantum fields with the matter. A generalization of the
boundary conditions method has been proposed by Symanzik
\cite{Symanzik'81}. In the framework of path integral formalism
he showed that presence of material boundaries (two dimensional
defects) in the system can be modeled with a surface term added to
the action functional. Such singular potentials with $\delta$-function profile
concentrated on the defect surface reproduce some simple
boundary conditions (namely Dirichlet and Neumann ones) in the
strong coupling limit. The additional action of the defect should
not violate basic principles  of the bulk model such as gauge
invariance (if applicable), locality and renormalizability.

The QFT systems with $\delta$-potentials are mostly investigated
for scalar fields. In \cite{Markov Pismak'05}--\cite{FMP 08} the
Symanzik approach was  for the first time used
to describe similar problems in complete quantum electrodynamics (QED), and
all $\delta$-potentials consistent with QED basic principles were
constructed.

It seems quite natural to try applying the same method for description
of interaction of quantum fields with bulk macroscopic
inhomogeneities (slabs, finite width mirrors, etc) and to study
Casimir effects in system of such a kind. There were different
attempts to quantize electrodynamics in presence of dielectric media
(i.e. volume inhomogeneities of special kind) see, for instance, \cite{Bordag'98},
\cite{Eberlein Robaschik 05}, none of them was truly successful.
The Symanzik's method was used to model the interaction of
quantum fields with bulk defects in a number of papers
(e.g. \cite{Bordag'95},\cite{Graham Jaffe}-%
\cite{Cavero-Pelaez Milton Wagner 05}, and others). However most of them
were devoted to study of a limiting procedure of transition from
a bulk potential of the defect to the surface $\delta$-potential as in
\cite{Graham Jaffe}.
%
On the other hand, results for the Casimir energy of a single
planar layer of finite width $\ell$ are contradictory. Thus, the
formulae presented recently in \cite{Fosco Lombardo Mazzitelli 08}
does not coincide with previous calculations made in
\cite{Bordag'95}. Moreover, the only attempt to calculate the
propagator in such system was undertaken in \cite{Aguiar 93} where
hardly any explicit formulae were after all presented.

Thus, one can see that the specificity of finite volume effects generated by
inhomogeneities in QFT has not been yet adequately explored.
Our work is dedicated to clarify the problem, and to solve existing controversy
within an accurate and unambiguous approach. We consider a model of massive
scalar field interacting with volume defect (finite width slab),
calculate the modified propagator of the field, the Casimir energy of the slab
and discuss its physical meaning.

\section{Statement of problem}
Let us consider a model of a real scalar field interacting with a
volume defect. In the simplest case such defect could be
considered as homogenous and isotropic infinite plane layer of the
thickness $\ell$, placed in the $x_1x_2$ plane. Generalizing the
Simanzik approach, we describe the interaction of quantum fields
with matter by introducing into the action of the model an
additional mass term which is non-zero only inside the defect
 \be
 \label{S}
    S=\frac12\int d^4x \(\phi(x) (-\partial^2_x+m^2) \phi(x)
        + \lambda \theta(\ell, x_3)\phi^2(x)\)
 \e
where $\partial_x^2=\partial^2/\partial
x_0^2+\ldots+\partial^2/\partial x_3^2$%
\footnote{We operate in Euclidian version of the theory which appears to be more
convenient for calculations.}).
The distribution function
$\theta(\ell, x_3)$ is equal to $1/\ell$ when $|x_3|<\ell/2$, and is
zero otherwise, in terms of the Heaviside step-function we can
write it as
$\theta(\ell, x_3)\equiv
[\theta(x_3+\ell/2)-\theta(x_3-\ell/2)]/\ell$. Such kind of potential
is also called patchwise (or piecewise) constant one.
In the framework of QFT it was considered for the first time in \cite{Bordag'95}, and later
in \cite{Feinberg Mann Revzen 99}-\cite{Fosco Lombardo Mazzitelli 08}.

To describe all physical properties of the systems it is
sufficient to calculate the generating functional for Green's
functions \be
    G[J]= N\int D\phi\, \exp\{-S[\phi]+J\phi\},
        \ N^{-1}=\int D\phi\, \exp\{-S_0[\phi]+J\phi\}
        \label{G(J)}
\e where $J$ is an external source,
$S_0(\phi)=S(\phi)|_{\lambda=0}$, and  normalization for the
generating functional we have chosen in such a way that
$G[0]|_{\lambda=0}=1$.

Introducing in (\ref{G(J)}) auxiliary fields $\psi$ defined in the
volume of the defect only, we can present the defect contribution to
$G[J]$ as
\be
 \exp\left\{-\frac{\lambda}{2\ell}\int d\vx\int_{-\ell/2}^{\ell/2} dx_3 \phi^2(x) \right\}
   =C \int D\psi \exp\left\{\int d\vx\int_{-\ell/2}^{\ell/2} dx_3\(-\frac{\psi^2}{2}
                                    +i\sqrt{\kappa}\psi\phi\)\right\}
\e
where $C$ is an appropriate normalization constant,
and $\kappa = {\lambda}/{\ell}$.

With help of projector onto the volume of
defect ${\cal O}=\theta(x_3+\ell/2)-\theta(x_3-\ell/2)$ acting as
$$
    \psi{\cal O}\phi\equiv \int d\vx\int_{-\ell/2}^{\ell/2} dx_3\psi\phi,
$$
we can perform functional integration over $\phi$, and
consequently over $\psi$. As a result we get
\be
G[J]= [{\rm Det}Q]^{-1/2}
    e^{\frac12J\hat S J}, \quad
\hat S=D-\kappa (D{\cal O}) Q^{-1}({\cal O} D),
    \label{G_fin}
\e
\be
Q=\textbf{1}+\kappa({\cal O} D{\cal O}).
 \label{Qx}
\e
Here the unity operator $\textbf{1}$, as well as the whole $Q$, is defined in the volume
of the defect only $(-\ell/2,\ell/2)\times{\mathbb R}^3$,
and $D=(-\partial^2+m^2)^{-1}$  is  the standard
propagator of free scalar field. We shall note here that the
outlook of (\ref{G_fin}) completely coincides with expression for
generating functional $G[J]$ in the case of delta-potential term
instead of patchwise constant one. It is also evident that a
straightforward generalization is possible for non-constant
$\kappa$($=\lambda/\ell$) with $\lambda$ depending on $x_3$.

In this paper we calculate explicitly both the modified propagator of
the system and its Casimir energy, and reveal their dependence on the
parameter $\lambda$ describing the material properties of the
homogeneous defect layer and its thickness~$\ell$.

\section{Calculation of the propagator}
To calculate the propagator $\hat{S}$  defined according to (\ref{G_fin})
let us first derive explicit formula for the operator $W\equiv Q^{-1}$.

For this purpose we first introduce the Fourier transformation
of the  coordinates parallel to the defect (i.e.  $x_0$, $x_1$, $x_2$).
Then for the propagator $D$ of the system without a defect one can write
$$
D(x)=\int\frac{d^3\vp}{(2\pi)^3}e^{i \vp \vx}
    \int\frac{dp_3}{(2\pi)} \frac{e^{i p_3 x_3}}{p_3^2+\vp^2+m^2},
$$
integrating over $p_3$ with help of the residue theorem we get
\be
D(x)
    =\int\frac{d^3\vp}{(2\pi)^3}e^{i \vp \vx}\, {\cal D}_{E^2}(x_3),
    \label{fourier}
    \qquad {\cal D}_{V}(x)\equiv\frac{e^{-\sqrt{V}|x|}}{2\sqrt{V}}
\ee
with $E=\sqrt{p^2+m^2}$.
Then we are able
to write the defining (operator) equation for $W$ as
\be
    W + \kappa {\cal D}_{E^2} W =1.
    \label{W}
\ee
By construction the mixed $\vp$-$x_3$ representation
of the free scalar propagator  ${\cal D}_{E^2}(x,y)\equiv{\cal D}_{E^2}(x-y)$
is the Green's function of the following ordinary differential operator
\be
\label{s0}
    K_{V}(x,y)=\left(-\frac{\partial^2}{\partial x^2}+V\right)\delta(x-y)
\ee
for $V=E^2$.
Multiplying both sides of (\ref{W}) with $K_{E^2}$
and using obvious relation $K_V-K_{V'}=V-V'$ we get
\begin{equation}
    K_\rho U = -\kappa
    \label{s1}
\end{equation}
where  $\rho\equiv \kappa+E^2$ and $U\equiv W-1$.

The general solution to this (inhomogeneous) operator equation can be written as a sum
of its partial solution and the general solution of its homogeneous version.
Then with help of ${\cal D}_{\rho}$ one writes for $U$
$$
    U(x,y)=-\kappa {\cal D}_{\rho}(x,y)
        +\alpha(y) e^{x \sqrt{\rho}}+\beta(y) e^{-x \sqrt{\rho}}.
$$
Here $\alpha$ and $\beta$ --- arbitrary functions on $y$. Imposing the symmetry condition
$U(x,y)=U(y,x)$ we derive that
$$
U(x,y)=-\kappa {\cal D}_{\rho}(x,y)
    +a e^{(x+y) \sqrt{\rho}}+b\left(e^{(x-y) \sqrt{\rho}}+e^{(y-x) \sqrt{\rho}}\right)
        +c e^{-(x+y) \sqrt{\rho}}
$$
where $a$, $b$ and $c$ are some constants now. Introducing
$W=1+U$ into (\ref{W}) one gets
\be
    U+\kappa {\cal D}_{E^2}(1+U)=0.
    \label{U_ident}
\ee
Requiring that this equation is an identity for all $x$ and
$y$ (we remind that $U\equiv U(x,y)$), we find for $a$, $b$ and $c$
\be
a=c=-\frac{\xi \kappa^2 e^{\ell \sqrt{\rho}}} {2 \sqrt{\rho}},\qquad
b=-\frac{\xi \kappa (E-\sqrt{\rho})^2}
    {2 \sqrt{\rho}},
           \label{abc}
\ee
$$
\xi=\frac1{e^{2\ell\sqrt\rho}(E+\sqrt\rho)^2-(E-\sqrt\rho)^2}.
$$

With help of these expressions we can finally
derive the explicit formula for the modified propagator
of the system. From the definitions of $\hat S$  and
$W$, and using (\ref{U_ident}) we can write that
\be
    \hat S =
        (1+U) {\cal D}_{E^2}.
    \label{S hat}
\ee
We divide the general expression of $\hat S\equiv\hat S (\vp,x_2,y_3)$ into four parts
according to the position of $x_3$, $y_3$ relative to the defect
\be
\hat S(\vp, x_3,y_3)=\left\{%
\begin{array}{ll}
    S_{--},         & x_3<-\ell/2,\ y_3<-\ell/2\\
    S_{-\circ},     & x_3<-\ell/2,\ y_3\in(-\ell/2,\ell/2) \\
    S_{-+},         & x_3<-\ell/2,\ y_3>\ell/2\\
    S_{\circ \circ}, & x_3\in(-\ell/2,\ell/2),\ y_3\in(-\ell/2,\ell/2) \\
\end{array}%
\right.
\ee
Other cases could be easily derived using the symmetry properties of the propagator.

Performing necessary integration according to (\ref{S hat}) we get
\begin{eqnarray}
  &S_{--}& = {\cal D}_{E^2}(x_3-y_3)
    +\frac{\xi\kappa e^{E \ell}(1-e^{2 \ell\sqrt\rho})}{2 E}e^{E (x_3+y_3)}
    \label{S fin}\\
  &S_{-\circ}& 
   = {\xi  e^{E(x_3+\ell/2)} e^{\ell\sqrt\rho}
        \((\sqrt\rho-E) e^{\sqrt\rho(y_3-\ell/2)}+(\sqrt\rho+E)e^{\sqrt\rho (\ell/2-y_3)}\)} \nonumber\\
  &S_{-+} &= 2 \xi  \sqrt\rho e^{(\sqrt\rho +E)\ell+E(x_3-y_3)}\nonumber\\
  &S_{\circ \circ} &= \frac{\xi e^{\ell\sqrt\rho}}{2 \sqrt\rho }
    \(
        2 \kappa \cosh[(x_3+y_3)\sqrt\rho]
        +e^{\sqrt\rho (|x_3-y_3|-\ell)} (E-\sqrt\rho)^2
        +e^{\sqrt\rho (\ell-|x_3-y_3|)} (E+\sqrt\rho)^2
    \)\nonumber
\end{eqnarray}
with $\xi$ defined in (\ref{abc}).

To the best of our knowledge the only attempt to calculate the propagator for such system
was presented in \cite{Aguiar 93} where its
final expression was given in terms of ``coefficients of scattering wave functions'' of
one-dimensional time-dependent Schrodinger equation. However, for the explicit formulae
for those coefficients the author refers yet to another paper \cite{Varro 90} (actually,
there is also a misprint in the reference number),
where the problem of
electrons scattering in a powerful laser field is considered and corresponding coefficients
are presented in the from of infinite series of Bessel functions. The result presented
in (\ref{S fin}) is in much simpler closed from, and it raises doubts of
correctness of calculations presented in \cite{Aguiar 93}, \cite{Varro 90}.

\section{The Casimir Energy}
It is well know that the Casimir energy density per unit area of the defect $S$
can be presented with the relation
\be
{\cal E}=-\frac{1}{TS}\ln G[0]
    =\frac1{2TS}\,\mbox{Tr}\ln[Q(x,y)].
    \label{E_TrLn}
\e
in the second equation we used (\ref{G_fin}).
For explicit calculations we first make the Fourier transformation
as in (\ref{fourier}). Then
\be
{\cal E}
    =\mu^{4-d}\,\int\frac{d^{d-1}\vp}{2(2\pi)^{d-1}}\mbox{Tr}\ln[Q(\vp; x_3,y_3)],
    \label{E_TrLn2}
\e
where we also introduced dimensional regularization to handle
UV-divergencies and an auxiliary normalization mass parameter $\mu$.

Using the definitions of $U$ and $Q$ we can express $\kappa$-derivative
of the integrand of (\ref{E_TrLn2}) in the following form
$$
    \partial_\kappa \ln Q={\cal D}_{E^2} W =-\frac{U}{\kappa}.
$$
Then for the energy density we get
\be
\label{res1}
{\cal E}
    =-\mu^{4-d}\,\int_0^\kappa \frac{d\kappa}{\kappa}
        \int\frac{d^{d-1}\vp}{2(2\pi)^{d-1}}\mbox{Tr}\, U.
\ee
We have chosen the lower limit of integration over $\kappa$
to satisfy the energy normalization condition ${\cal E}|_{\kappa=0}=0$.
As we show below the integral is convergent at $\kappa=0$.

The trace of the integral operator $U$ is straightforward
\be
\mbox{Tr}\, U\equiv \int_{-\ell/2}^{\ell/2} dx\, U(x,x)
    =2 b\ell+\frac{4 a \sinh(\ell\sqrt\rho)-\ell\kappa}{2\sqrt\rho}
    \label{Tr U}
\ee
where we already used that $a=c$. Using $a$ and $b$
given in (\ref{abc}), one easily notes that
$\mbox{Tr} U\sim - \ell \kappa/(2 E)$ when $\kappa\to0$, thus supporting the above statement.

Next, putting (\ref{Tr U}) into (\ref{res1})
we can compare our result with previous calculations performed in \cite{Bordag'95},
and also recently rederived in \cite{Vass}.
Instead of explicit $\kappa$-integration in (\ref{res1}), we can differentiate the
above mentioned result by Bordag with respect to $\kappa$ to see immediately that
it coincides explicitly with integrand of (\ref{res1}). Thus, we write for the energy
\be
{\cal E}=\mu^{4-d}\int \frac{d^{d-1}\vp}{2 (2\pi)^{d-1}}
    \ln\[
        \frac{e^{- \ell E}}{4 E \sqrt\rho}
            \(e^{ \ell\sqrt\rho }(E+\sqrt\rho)^2-e^{- \ell\sqrt\rho }(E-\sqrt\rho)^2\)
    \]
    \label{E}
\ee

To extract the UV divergencies in $d=4$,
let's consider those contributions in ${\cal E}$ (\ref{E}) that do not converge
while integrated over $p$. We have
$$
\ln\[
        \frac{e^{- \ell E}}{4 E \sqrt\rho}
            \(e^{ \ell\sqrt\rho }(E+\sqrt\rho)^2-e^{- \ell\sqrt\rho }(E-\sqrt\rho)^2\)
    \]=
 \frac{\lambda}{2 E}-\frac{\lambda^2}{8 \ell E^3}+
O\(\frac{1}{E^4}\),
$$
Hence, within dimensional regularization the energy  can  be represented as follows
$$
    {\cal E}= {\cal E}_{fin} + {\cal E}_{div},
$$
where
$$
{\cal E}_{fin}= \frac1{2 \pi^2}\int_0^\infty \Xi(p) p^2 dp ,
$$
\be
\Xi(p)\equiv
    \ln\[
        \frac{e^{-2 \ell E}}{4 E \sqrt\rho}
            \(e^{2\ell\sqrt\rho }(E+\sqrt\rho)^2-e^{-2\ell\sqrt\rho }(E-\sqrt\rho)^2\)
    \]
    -\frac{\lambda}{4E}\(1-\frac{\lambda}{4 \ell E^2}\),
\label{E_fin}
\ee
$$
{\cal E}_{div}=\frac{\lambda\mu^{4-d}}{2 (2 \pi)^{d-1}}
    \int \frac{d^{d-1} p}{4E}\(1-\frac{\lambda}{4 \ell E^2}\).
$$
The first item  ${\cal E}_{fin}$ is finite and we removed
regularization,  while ${\cal E}_{div}$ is divergent but trivially
depends on the parameters of the theory and auxiliary parameter
$\mu$.
We add now to the action of the model a field-independent counter-term $\delta
S$ of the form $\delta S = f  + g \ell^{-1}$, with bare parameters
$f$ and $g$ (of mass dimensions two and one correspondingly). It
allow us to choose these parameters in such way that the
renormalized Casimir energy ${\cal E}_{r}$ defined by the full
action $S+\delta S$ and considered as the function of renormalized
parameters appears to be finite both in regularized theory, and
also after the removing of regularization.

Thus, for the renormalized Casimir energy we obtain the following result
\be
    {\cal E}_r= {\cal E}_{fin} + f_r + \frac{g_r }{\ell}
    \label{E_ren}
\e
where finite  parameters $f_r$, $g_r$ must be determined with appropriate experiments.

The Casimir pressure on the slab  is then
$$
p=-\frac{\partial {\cal E}_r}{\partial \ell}=
    -\frac{\partial {\cal E}_{fin}}{\partial \ell}+   \frac{g_r}{\ell^{2}}.
$$
Taking into account the definition of distribution function
$\theta(\ell, x_3)$ one can say that the derivative is taken here
on condition that the amount of matter (effectively described by
the defect) in the slab is fixed: $\int dx_3 \theta(\ell, x_3)=1$.
Alternatively, one can consider the density of the matter to be
fixed and calculate the pressure under this condition. Then the
distribution function has a different normalization condition
$\int dx_3 \theta(\ell, x_3)=\ell$, which is equivalent to the
mere change of variables  $\lambda\to\ell \tilde\lambda$ in the
formula (\ref{E_fin}).

\section{Conclusion}

We constructed QFT model of the scalar field interacting with the bulk defect
concentrated within a slab of finite width $\ell$.
The propagator and the vacuum determinant (Casimir energy) were calculated.
The later one coincides with results obtained in \cite{Bordag'95}, \cite{Vass}
within a different approach, while the explicit formula for the propagator is
given for the first time.
The Casimir energy is UV divergent and for its regularization we
applied dimensional regularization. It allowed us to extract the finite part
and to construct the counter-terms. The renormalization procedure
requires generally two normalization conditions to fix the values
of the counter-terms with the appropriate experiments.
It is shown that the Casimir pressure in the system can be calculated
in two different ways: for fixed density of matter and for fixed amount of matter of the slab.

Similar problems were considered recently in \cite{Fosco Lombardo
Mazzitelli 08} in the framework of massless scalar field
interacting with a slab (mirror) of general profile. However, the
massless limit of our result for the Casimir energy of a single
slab (\ref{E}) differs from one obtained in \cite{Fosco
Lombardo Mazzitelli 08}, Eq. (68) for the case of `piecewise constant'
profile (equivalent to our case).
%
As a validity check we appeal to the general perturbation theory.
Decomposing the generating functional
$G(J)$ (\ref{G(J)}) in a perturbation series in $\lambda$, one finds that for
the massive theory $G(0)$ is analytical at $\lambda=0$ with irrelevant
(geometry independent) linear term. However, it is
evident that naive perturbation expansion fails for the limit
$m\to0$, alerting us of non-analyticity of the vacuum energy at
$\lambda=0$. Expanding Eq. (68) of \cite{Fosco Lombardo Mazzitelli 08}
in a power series in $\lambda$ one can easily see that it is
perfectly analytical with non vanishing linear term,
and thus does not comply with this general
argument. At the same time both the massive
and massless limits of our result (\ref{res1}),
which is equivalent to (\ref{E}) derived
independently by two other groups, does posses the required
(non-) analyticity properties.


In our work we considered a model of interaction of quantum scalar field
with material slab assuming $\lambda>0$. One must note that with a simple
redefinition of the parameters of the system under consideration
(i.e. $\lambda=-2 m^2 \ell$)
one can calculate the Casimir energy of two semi-infinite slabs
separated by a vacuum gap and interacting through a massless scalar
field. Similar problem in the framework of quantum statistical
physics was first solved by Lifshitz, \cite{Lifshitz'56}.
Comparison with Lifshitz formula, and further generalization of
the method proposed in this paper to the case of QED is the scope
of our future work.

\section*{Acknowledgement}
Authors are grateful to Prof. Vassilevich for attracting our attention to some literature on the
issue, and to Prof. Mazzitelli for his interest in our research.

V.N. Markov and Yu.M. Pismak are also grateful to Russian Foundation of Basic Research
for financial support (RFRB grant $07$--$01$--$00692$).

\end{document}